\documentclass[a4paper, twocolumn]{revtex4} 
\usepackage{color}
\usepackage{graphics}
\usepackage{epsfig}
\begin{document}
	
	\title{Evolution of spin freezing transition and structural, magnetic phase diagram of Dy$_{2-\textit{x}}$La$_\textit{x}$Zr$_{2}$O$_{7}$ (0 $\leq$ $\textit{x}$ $\leq$ 2.0)}
	
	\author{Sheetal and C. S. Yadav*}
	\affiliation{School of Basic Sciences, Indian Institute of Technology Mandi, Mandi-175075 (H.P.), India}
	
	\begin{abstract}
		Dy$_{2}$Zr$_{2}$O$_{7}$ a disordered pyrochlore system, exhibits the spin freezing behavior under the application of the magnetic field. We have performed detailed magnetic studies of Dy$_{2-\textit{x}}$La$_\textit{x}$Zr$_{2}$O$_{7}$ to understand the evolution of the magnetic spin freezing in the system. Our studies suggest the stabilization of the pyrochlore phase with the substitution of non-magnetic La along with the biphasic mixture of  fluorite and pyrochlore phases for the intermediate compositions. We observed that the spin freezing (\textit{T}$_{f}$ $\sim$ 17 K) at higher La compositions (1.5 $\leq$ $\textit{x}$ $\leq$ 1.99) is similar to the field-induced spin freezing for low La compositions (0 $\leq$  $\textit{x}$ $\leq$ 0.5) and the well-known spin ice systems Dy$_{2}$Ti$_{2}$O$_{7}$ and Ho$_{2}$Ti$_{2}$O$_{7}$. The low-temperature magnetic state for higher La compositions (1.5 $\leq$ $\textit{x}$ $\leq$ 1.99) culminates into a spin-glass like state below 6 K. Cole-Cole plot and Casimir-du Pr$\acute{e}$ fit shows the narrow distribution of spin relaxation time in these compounds.
		
	\end{abstract}
	
	\maketitle
	
		The geometrical frustration is one of the key aspects of the cubic pyrochlore oxide A$_{2}$B$_{2}$O$_{7}$ owing to the distorted spin geometry, and negligible structural disorder \cite{villain1979insulating, binder1986spin}. In pyrochlore structure, A and B sites form a network of corner-sharing tetrahedra, a quintessential framework for a geometrically frustrated magnet. These materials display a variety of exotic phases like spin ice (Dy$_{2}$Ti$_{2}$O$_{7}$ and Ho$_{2}$Ti$_{2}$O$_{7}$), spin liquid (Tb$_{2}$Ti$_{2}$O$_{7}$), spin glass (Y$_{2}$Mo$_{2}$O$_{7}$), order by disorder (Er$_{2}$Ti$_{2}$O$_{7}$), Kondo effect (Pr$_{2}$Ir$_{2}$O$_{7}$), unconventional anomalous Hall effect (Nd$_{2}$Mo$_{2}$O$_{7}$), superconductivity (Cd$_{2}$Re$_{2}$O$_{7}$), etc. \cite{snyder2004low,ehlers2002dynamical,enjalran2004spin,greedan1986spin,petrenko2013low,nakatsuji2006metallic,taguchi2003magnetic,jin2001superconductivity}.
	
	The magnetic spin ice systems have offered an interesting physics with the signature of magnetic monopole like excitation at low temperature \cite{toews2018disorder,morris2009dirac,jaubert2011magnetic}. Apart from the spin-ice behavior below $\sim$ 2 K, Dy$_{2}$Ti$_{2}$O$_{7}$ also shows a strong frequency dependent spin-freezing at $\sim$ 16 K. However no such feature is seen in the spin-ice Ho$_{2}$Ti$_{2}$O$_{7}$ which distinguishes the spin dynamics of (Ho/Dy)$_{2}$Ti$_{2}$O$_{7}$ \cite{bramwell2001spin,harris1997geometrical}. Although the Dy$_{2}$Ti$_{2}$O$_{7}$ and Ho$_{2}$Ti$_{2}$O$_{7}$ systems are extensively studied for spin ice properties, a complete understanding of the phenomenon is still elusive \cite{castelnovo2008magnetic,fennell2009magnetic}. The spin freezing behavior of Dy$_{2}$Ti$_{2}$O$_{7}$ is very different from the freezing in spin glasses. Unlike the typical spin glass systems the freezing transition in Dy$_{2}$Ti$_{2}$O$_{7}$ enhances with the application of magnetic field and give rise to unusual glassiness \cite{snyder2004low}. Theoretically, an ideal spin-glass behavior is predicted on the pyrochlore lattice of infinitely large dimensions \cite{kurchan2012exact,saunders2007spin}. For the finite dimensional system glass phase is established in the systems that exhibit quenched disorder. \cite{binder1986spin}. However, some of the pyrochlore magnets show spin glass transition, even in the absence of quenched disorder, which is indistinguishable to the typical spin glass systems \cite{taniguchi2009spin,gardner1999glassy,reimers1991short}. 
	
	Among these pyrochlore systems, Dy$_{2}$Zr$_{2}$O$_{7}$ (DZO) is quite fascinating, as it shows the emergence of the magnetic field induced spin freezing near 10 K and possess the magnetic entropy of R[ln2 - (1/2)ln(3/2)] in the presence of magnetic field of 5 kOe, which is same as the spin ice Dy$_{2}$Ti$_{2}$O$_{7}$ \cite{devi2020emergence}. The DZO crystallizes in the  pyrochlore phase and does not exhibit any magnetic ordering down to 40 mK \cite{ramon2019absence}. Further partial substitution of Dy by La (up to 15$\%$) stabilizes the pyrochlore phase and spin freezing is observed at a lower magnetic field in comparison to DZO \cite{devi2020emergence}. 
	
	The stability of the pyrochlore structure and the magnetic rare-earth atom (Dy, Ho, etc.) plays a very important role in determining the low-temperature magnetic spin ice ground state of these systems. In this light, we have studied the structural and magnetic phase diagram of Dy$_{2-\textit{x}}$La$_\textit{x}$Zr$_{2}$O$_{7}$ system. We extend our previous studies of diluted  pyrochlore zirconate to a much broader range of dilutions to 0 $\leq$ \textit{x} $\leq$ 2.0. In doing so, we observed the evolution of structure from disordered pyrochlore (0 $\leq$ $\textit{x}$ $\leq$ 0.5) to stable pyrochlore phase (1.5 $\leq$ $\textit{x}$ $\leq$ 2.0) through the biphasic mixture of these phases for 0.5 $<$ $\textit{x}$ $<$ 1.5. The compounds with disordered pyrochlore structures do not show spin freezing in the absence of a magnetic field. In contrast, the stable pyrochlore compounds exhibit Dy$_{2}$Ti$_{2}$O$_{7}$ like spin freezing at 17.5 K. The low-temperature phase of 1.5 $\leq$ $\textit{x}$ $\leq$ 1.99 compositions show glass-like behavior below 6 K.   
	
		DZO exhibits disordered pyrochlore structure (space group: Fd$\bar{3}$$\textit{m}$) with the remnants of pyrochlore phase \cite{devi2020emergence,ramon2019absence,mandal2006preparation,sayed2011sm2,glerup2001structural}. The XRD patterns of Dy$_{2-\textit{x}}$La$_\textit{x}$Zr$_{2}$O$_{7}$ reveal that the compositions corresponding to 0 $\leq$ $\textit{x}$ $\leq$ 0.5, adopt disordered fluorite structure. For 0.5 $<$ $\textit{x}$ $<$ 1.5, a few super-structure peaks corresponding to pyrochlore structure appear at 2$\textit{$\theta$}$ = 14$^{o}$ (\textit{111}), 27$^{o}$ (\textit{311}), 37$^{o}$ (\textit{331}), 45$^{o}$ (\textit{511}), etc. and the main peaks belonging to pyrochlore structure get split. The xrd data in this range was nicely fitted with the biphasic mixture of disordered pyrochlore and clean pyrochlore lattice structures (see supplementary Fig. S3). The Rietveld refinement for 1.5 $\leq$ $\textit{x}$ $\leq$ 2.0 confirms the clean face-centered-cubic pyrochlore structure with the observation of super-structure peaks \cite{hagiwara2017crystal,rodriguez1990fullprof}. This is also anticipated from the closer value of ionic radius ratio (\textit{r}$_{A}$/\textit{r}$_{B}$) and position of oxygen atom x(O) in the pyrochlore structure (Fig. 1). For a perfect pyrochlore structure the \textit{r}$_{A}$/\textit{r}$_{B}$ ranges between 1.48 to 1.78 and ideal value of parameter x(O) $\sim$ 0.3125 \cite{pal2018high,kumar2019spin}. The deviation from this value, modifies the octahedral (BO$_{6}$) and cubic (AO$_{8}$) symmetry and results in the change in crytal-field level splitting. For $\textit{x}$ $\leq$ 0.5, the \textit{r}$_{A}$/\textit{r}$_{B}$ ratio is less than the expected range of the pyrochlores and the parameter x(O) is close to $\sim$ 0.362 which indicates the formation of disordered fluorite/pyrochlore structure. Here, with $\textit{x}$ $\geq$ 1.5, the \textit{r}$_{A}$/\textit{r}$_{B}$ (where \textit{r}$_{A}$ = 1.027/1.16 $\AA$ for Dy$^{3+}$/La$^{3+}$ and \textit{r}$_{B}$ = 0.72 $\AA$ for Zr$^{4+}$) found to be in the pyrochlore regime (Fig. 1a). The x(O) evolves continuously towards pyrochlore regime with La substitution (Fig. 1b) and supports the stabilization of pyrochlore structure for higher La concentrations. The transformation from disordered pyrochlore to ordered pyrochlore structure along with the positional parameter x(O) would presumably enhanced the crystal field spacing about A site. A systematic shift of peak position towards lower angle on La incorporation indicates the increase of unit cell parameter (see supplementary Fig. S1). 	
	
	\begin{figure}[ht]
	\begin{center}
		\includegraphics[width=9cm,height=7cm]{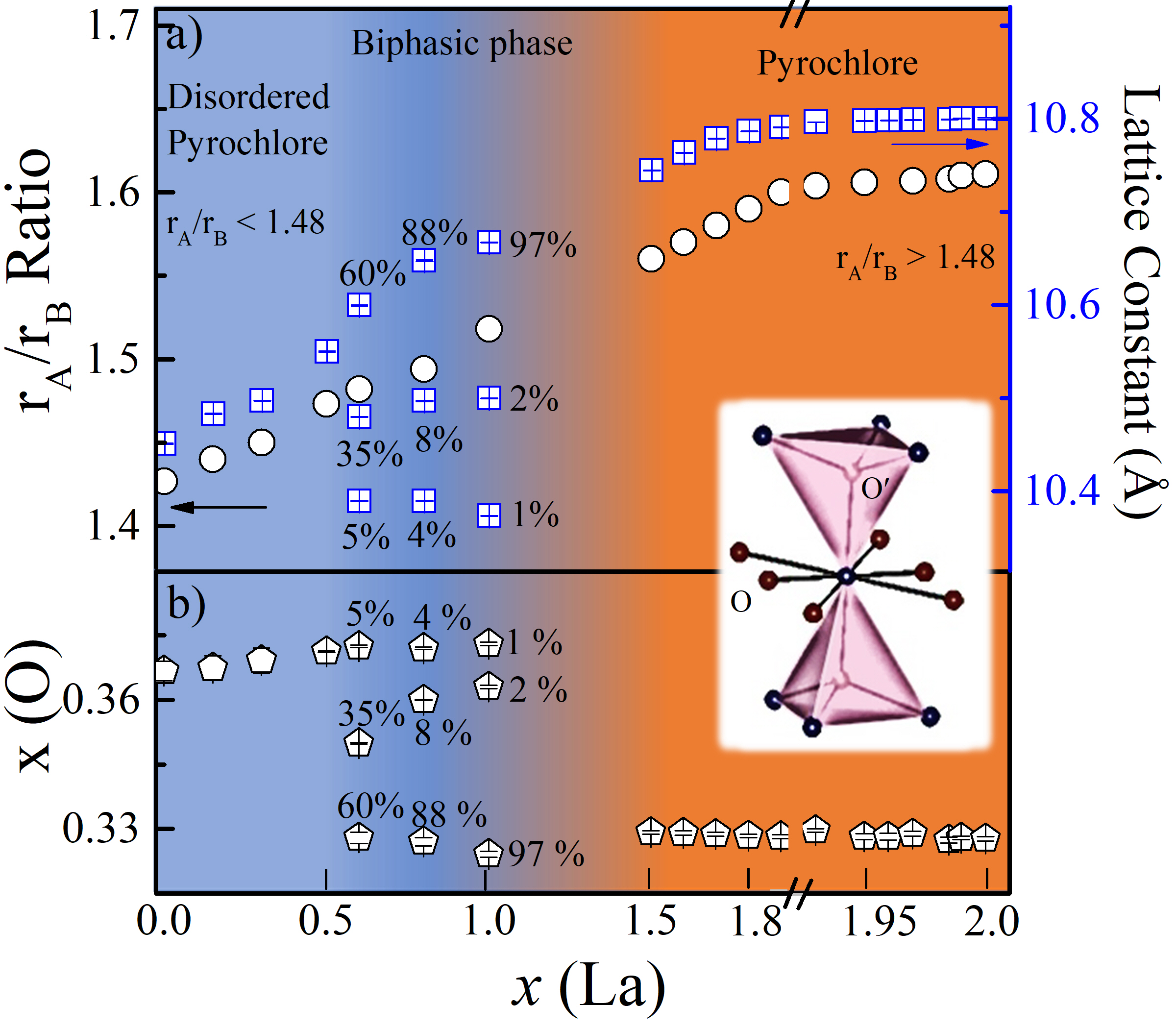}
		\vspace{-10pt}
		\caption{ (a) Variation in Radius ratio (\textit{r}$_{A}$/\textit{r}$_{B}$) (Left) and lattice constant (Right) as a function of La substitution for Dy$_{2-\textit{x}}$La$_\textit{x}$Zr$_{2}$O$_{7}$; (0 $\leq$ $\textit{x}$ $\leq$ 2). We have extended the scale between \textit{x} = 1.9 - 2.0 for the better clarity. (b) Variation in the position of x(O) parameter of oxygen atom as a function of La concentration. The percentage indicates the contribution from each phase in the refinement.}
		\vspace{-20pt}
	\end{center}
\end{figure}

	\begin{figure}[ht]
	\begin{center}
		\includegraphics[width=8cm,height=7cm]{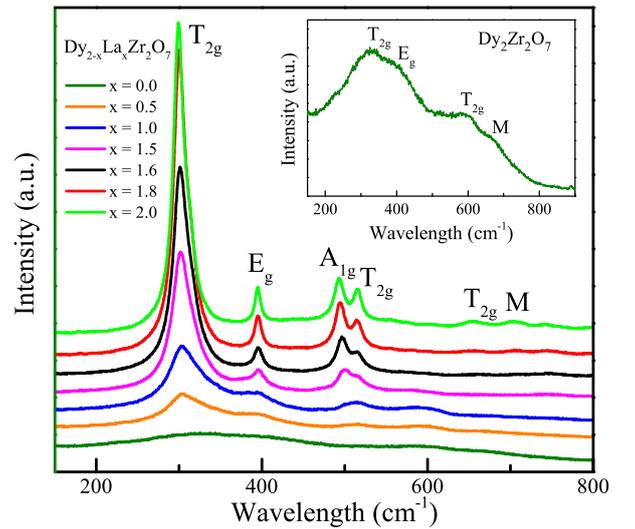}
		\caption{Room temperature Raman spectra of Dy$_{2-\textit{x}}$La$_\textit{x}$Zr$_{2}$O$_{7}$; (0 $\leq$ $\textit{x}$ $\leq$ 2) showing the presence of Raman modes corresponding to pyrochlore structure. Inset: Raman spectra of Dy$_{2}$Zr$_{2}$O$_{7}$.}
		\vspace{-15pt}
	\end{center}
\end{figure}
	
	Raman spectra of a pyrochlore structure consist of six Raman active modes; $\textit{A}$$_{1g}$, $\textit{E}$$_{g}$, and four $\textit{T}$$_{2g}$ modes; corresponding to the vibrations of $<$A-O$>$ and $<$B-O$>$ bonds \cite{han2015electron,hasegawa2010raman}. The DZO shows no mode (Fig. 2) at 466 cm$^{-1}$ corresponding to the fluorite structure but exhibits the pyrochlore type structural ordering with the presence of weak modes of pyrochlore structure \cite{hozoi2014longer, devi2020emergence}. The modes become more intense upon La substitution. The Raman spectra of the biphasic compositions (0.5 $<$ $\textit{x}$ $<$ 1.5) also do not exhibit any extra mode other than the pyrochlore structure and thus indicate the absence of any impurity phase in these compounds. The Raman spectra of Dy$_{2-\textit{x}}$La$_\textit{x}$Zr$_{2}$O$_{7}$ for the compositions $\textit{x}$ $\geq$ 1.5 shows sharper and well pronounced peaks (at 301 cm$^{-1}$ ($\textit{T}$$_{2g}$), 395 cm$^{-1}$ ($\textit{E}$$_{g}$), 493 cm$^{-1}$ ($\textit{A}$$_{1g}$), 514 cm$^{-1}$ ($\textit{T}$$_{2g}$) and 651 cm$^{-1}$ ($\textit{T}$$_{2g}$) than for $\textit{x}$ $\leq$ 0.5. This is possibly due to the increase in structural symmetry which is consistent with \textit{r}$_{A}$/\textit{r}$_{B}$ ratio for La substituted DZO. Additionally, an extra mode at 701 cm$^{-1}$ (marked by M) is observed in all the substituted compounds which was observed around 650 cm$^{-1}$ in the parent compound and attributed to the vibrations of the ZrO$_{6}$ octahedra \cite{glerup2001structural,turner2017lanthanide}. 
	
\begin{figure}[ht]
	\begin{center}
		\includegraphics[scale=0.35]{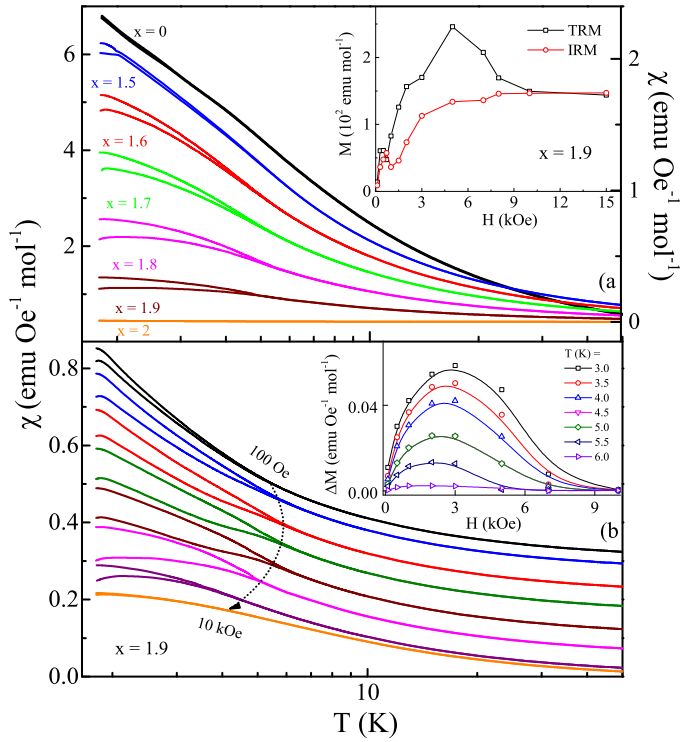}
		\vspace{-10pt}
		\caption{(a) Dc magnetic susceptibility \textit{$\chi$}$_{dc}$ (ZFC and FC) versus \textit{T} at \textit{H} = 100 Oe for Dy$_{2-\textit{x}}$La$_\textit{x}$Zr$_{2}$O$_{7}$ (0 $\leq$ $\textit{x}$ $\leq$ 2). Inset: Thermo and isothermo remnant magnetization as a function of field at $\textit{T}$ = 1.8 K for Dy$_{0.1}$La$_{1.9}$Zr$_{2}$O$_{7}$. (b) Dc magnetic susceptibility $\chi_{dc}$ (ZFC and FC) versus \textit{T} at various dc applied field for Dy$_{0.1}$La$_{1.9}$Zr$_{2}$O$_{7}$ (data is shifted along y-axis for better visibility of $\textit{T}$$_{irr}$). Inset: Field response of $\Delta M$ = M$_{FC}$ - M$_{ZFC}$ of Dy$_{0.1}$La$_{1.9}$Zr$_{2}$O$_{7}$ at various temperature.}
		\vspace{-10pt}
	\end{center}
\end{figure}
	
We measured the magnetization of the compounds at a low field (100 Oe) using the zero-field-cooled (ZFC) and field-cooled (FC) protocols. In the previous study \cite{devi2020emergence}, we have shown that the compounds having low La composition in Dy$_{2-\textit{x}}$La$_\textit{x}$Zr$_{2}$O$_{7}$; \textit{x} = 0.0, 0.15, 0.3 exhibit paramagnet-like behavior down to 1.8 K. This behavior sustains up to the La substitution of \textit{x} = 1.0, but the dc magnetic susceptibility of \textit{x} = 1.5 shows bifurcation of ZFC and FC curves at \textit{T}$_{irr}$ $\sim$ 4 K (Fig. 3a). The FC magnetization is completely reversible, while the ZFC magnetization is irreversible at a fixed magnetic field. The \textit{T}$_{irr}$ (which is taken at the bifurcation point between the ZFC and FC magnetization) shifts towards higher temperatures for further increase in La concentration (shown in Fig. S6a) and disappeared with the complete replacement of Dy with La (\textit{x} = 2.0). The absence of \textit{T}$_{irr}$ in the end compound (La$_{2}$Zr$_{2}$O$_{7}$ : [Xe]5d$^{0}$6s$^{0}$) is obvious. The splitting in the ZFC and FC curves can be taken as an indication of the spin-glass phase. In conventional spin-glass systems the glass state is quenched on applying the sufficiently strong field, and the temperature at which the glass state appears decreases on increasing the magnetic field \cite{mydosh1993spin}. As shown in Fig. 3b, the bifurcation point \textit{T}$_{irr}$ slightly increases with the applied field up to 3 kOe and then decreases to low temperature on a further increase of field and disappear at 10 kOe. We have plotted the field response of \textit{$\Delta$M} =  \textit{M$_{FC}$} - \textit{M$_{ZFC}$} of Dy$_{0.1}$La$_{1.9}$Zr$_{2}$O$_{7}$ at various temperature in the inset of Fig. 3b. It is clearly seen that similar to the field dependence of \textit{T}$_{irr}$ the difference (\textit{$\Delta$M}) between the two data sets increases with the applied field up to 3 kOe and no difference between the FC and ZFC magnetization is observed at 10 kOe. Furthermore, the shifting of \textit{T}$_{irr}$ towards low temperature for \textit{H} $\geq$ 3 kOe and reduction in the absolute value of \textit{$\chi$}$_{dc}$ under the application of high magnetic field indicates the frozen spin-glass-like state below \textit{T}$_{irr}$. Spin glass is a metastable state which arises due to random site distribution and random exchange and leads to a multi-degenerate ground state \cite{mydosh1993spin, binder1986spin,huang1985some}. In pyrochlores, glassiness occurs possibly due to the competing interactions, which transform to the glass/frozen state on cooling without any long-range magnetic order. The substitution of non-magnetic La at the magnetic Dy site is bound to affect the magnetic interactions present in the system. The replacement of 3/4$^{th}$ of Dy atoms (\textit{x} = 1.5) reduces the overall magnetic interactions, leading to the establishment of spin-glass-like state. This further evident from MH isotherm, where La substitution leads to the 6 - 8$\%$ change in the spin anisotropy in comparison to the Dy$_{2}$Ti$_{2}$O$_{7}$ system.\\
The irreversibility of spin state at low-temperature was further confirmed by examining the isothermal remnant magnetization (IRM) and thermoremnant magnetization (TRM). The IRM and TRM curves shown in the inset of Fig. 3a are consistent with the spin-glass picture \cite{snyder2004low}. The difference between the magnetization curve below 10 kOe points towards the retained memory of its aged value under the same final condition. This shows that a sufficiently high field can destroy this memory and the behavior is equivalence of ZFC and FC data taken in the same field. Notably, both the magnetization curves merge at a field 10 kOe. Surprisingly, we do not observe any frequency dependence in the ac susceptibility measurement below \textit{T}$_{irr}$, which raises some doubt in the magnetic state. Thus, these observations point towards the formation of spin-glass-like state instead of a clean spin glass state. Considering the possibility of long-range cooperative interactions between rare-earth ions, systems often do not get into a clean magnetic state. Some additional local probe measurements are required for a clear understanding. The Curie Weiss fitting of dc susceptibility data (measured at \textit{H} = 100 Oe) in the temperature range 30 - 300 K shows the dominance of antiferromagnetic interactions in these systems. The obtained parameters are listed in Supplementary Table II. 
	
	\begin{figure}[ht]
	\begin{center}
		\includegraphics[scale= 0.35]{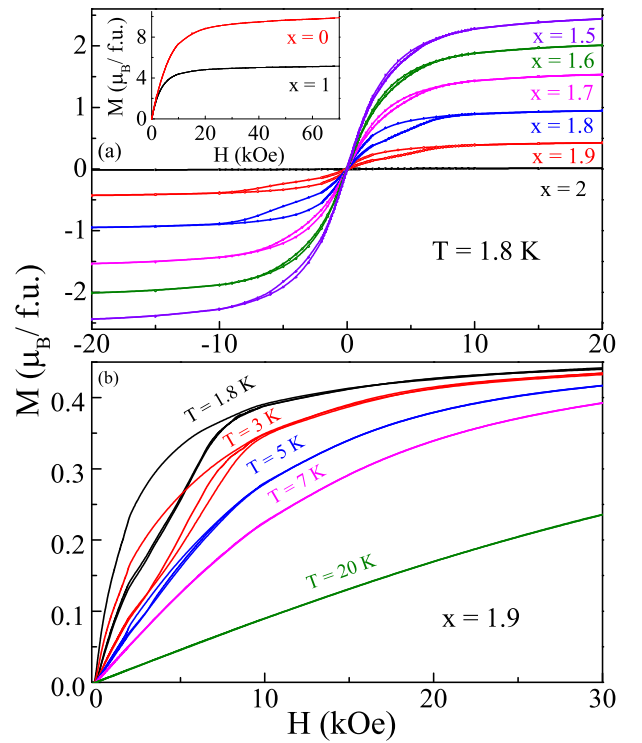}
		\vspace{-10pt}
		\caption{ (a) Isothermal Magnetization M(H) of Dy$_{2-\textit{x}}$La$_\textit{x}$Zr$_{2}$O$_{7}$; $\textit{x}$ = 1.5, 1.6, 1.7, 1.8 and 1.9 at 1.8 K. Inset shows the isothermal magnetization at 1.8 K for \textit{x} = 0, 1. (b) Field dependence of magnetization for Dy$_{0.1} $La$_{1.9}$Zr$_{2}$O$_{7}$ at \textit{T} = 1.8, 3, 5, 7 and 20 K.}
		\vspace{-10pt}
	\end{center}
\end{figure}

Figure 4a shows the field response of magnetization at 1.8 K. The magnetization isotherm for DZO (inset of fig. 4a) increases up to 20 kOe and remains saturated up to the maximum applied field of 70 kOe. The saturation moment (\textit{M}$_{s}$) for DZO (4.8 $\mu_{B}$/Dy) is close to the Dy$_{2}$Ti$_{2}$O$_{7}$ (5 $\mu_{B}$/Dy) but half of the theoretically expected value (10.64 $\mu_{B}$/Dy). This indicates strong crystal-field induced anisotropy in the system, which would possibly changed on dilution \cite{anand2015investigations,fukazawa2002magnetic}. The substitution of La up to \textit{x} $\leq$ 0.3 exhibits similar behavior as that of DZO \cite{devi2020emergence}. However, for the higher concentration of La (\textit{x} $\geq$ 1.5) magnetization isotherms show a large hysteresis below 7 K and get saturated at the lower field compared to DZO. The obtained values of saturation magnetization for La substituted compounds are nearly half of the free-ion value and vary only 6 - 8$\%$ to the \textit{M}$_{s}$ of Dy$_{2}$Ti$_{2}$O$_{7}$. These results are in line with crystal field-induced anisotropy in La substituted compounds that do not affect the magnetic moment to a greater extent. The effect of La substitution on anisotropy is measurably small and comparable to the dilution of Dy site with Ca, and quite distinct from the Y substitution. \cite{anand2015investigations,snyder2004quantum,lin2014nonmonotonic}. \\
The irreversibility seen in the dc magnetization (Fig. 3) around 6 K is evident from the magnetic isotherm curve. As shown in Fig. 4b, the magnetic hysteresis for Dy$_{0.1}$La$_{1.9}$Zr$_{2}$O$_{7}$ suppressed at 7 K, similar to the dc magnetization where ZFC and FC curves overlap at this temperature. Interestingly, no appreciable hysteresis is found at \textit{H} $\sim$ 10 kOe, at which the bifurcation between the ZFC-FC and IRM-TRM curves is lost. 
	
		\begin{figure}[ht]
	\begin{center}
		\vspace{-10pt}
		\includegraphics[width=9cm,height=12cm]{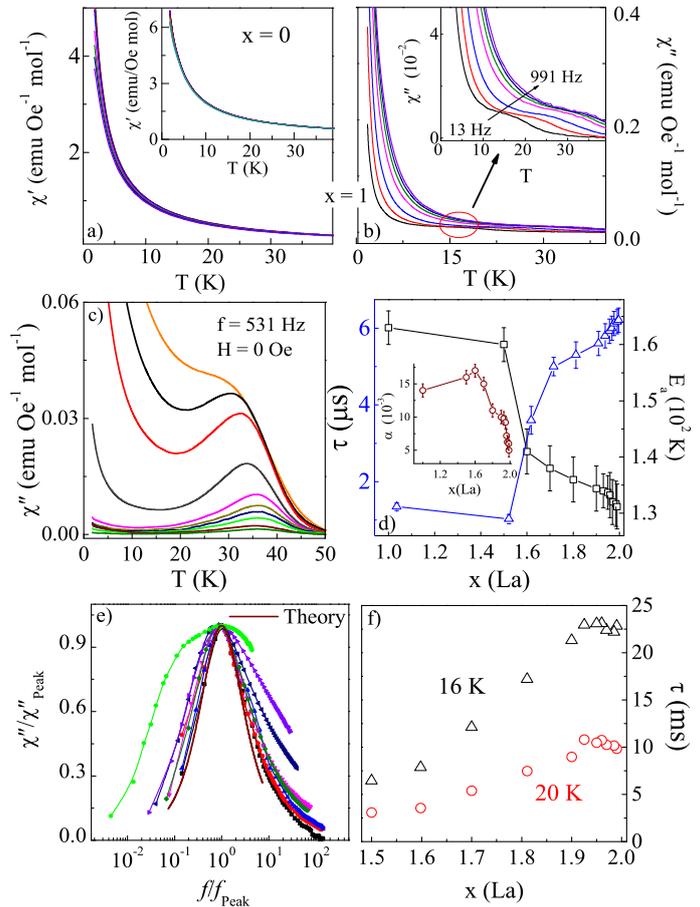}
		\caption{(a, b, c) The real \textit{$\chi$}$^{\prime}$ and imaginary \textit{$\chi$}$^{\prime\prime}$ parts of ac susceptibility for Dy$_{2-\textit{x}}$La$_\textit{x}$Zr$_{2}$O$_{7}$ (0 $\leq$ $\textit{x}$ $\leq$ 1.99) measured at H$_{dc}$ = 0 Oe for frequencies ranges between \textit{f} = 13 - 991 Hz. (d) Variation of E$_{a}$ and relaxation time $\textit{$\tau$}$  for 1 $\leq$ $\textit{x}$ $\leq$ 1.99. Inset shows the variation in Cole-Cole parameter $\alpha$. (e) Casimir-du Pr$\acute{e}$ relation fit for ac susceptibility (f) Relaxation time \textit{$\tau$} = 1/2$\pi$\textit{f} vs \textit{x} determined from the frequency of \textit{$\chi$}$^{\prime\prime}$ for 1.5 $\leq$ \textit{x} $\leq$ 1.99.}
	\end{center}
	\vspace{-15pt}
\end{figure}
	
	The ac susceptibility ($\chi_{ac}$) of Dy$_{2-\textit{x}}$La$_\textit{x}$Zr$_{2}$O$_{7}$ (0 $\leq$ \textit{x} $<$ 2) are shown in Fig. 5, measured to investigate the effect of La substitution on the spin dynamics of the system. The pyrochlore compound Dy$_{2}$Ti$_{2}$O$_{7}$ shows the high-temperature spin freezing anomaly at 16 K and spin ice ground state at low-temperature \cite{snyder2001spin}.  The replacement of Ti site with isovalent but comparatively larger Zr atom disorder the crystal structure and leads to a very dynamic ground state down to 40 mK \cite{ramon2019absence}. However, the disorder induced by the Zr atom (as discussed by Ramon \textit{et al.} \cite{ramon2019absence}) is significantly suppressed by applying the magnetic field and shows the re-entrance of Dy$_{2}$Ti$_{2}$O$_{7}$ like magnetic state at 10 K with the presence of non-zero residual entropy \cite{devi2020emergence} at 5 kOe. The compounds with La substitution of \textit{x} $\leq$ 0.5 exhibits the paramagnet-like behavior similar to the parent compound in real ($\chi^{\prime}$) and imaginary ($\chi^{\prime\prime}$) parts of \textit{$\chi$}$_{ac}$, measured at zero dc field.  Figs. 5a and 5b show the temperature variation of \textit{$\chi$}$^{\prime}$ and \textit{$\chi$}$^{\prime\prime}$ for frequency \textit{f} = 10 - 1000 Hz at zero dc applied field for DyLaZr$_{2}$O$_{7}$ (i.e. \textit{x} = 1.0). As seen, \textit{$\chi$}$_{ac}$ remains almost featureless up to this concentration, except for the development of weak features near 15 K. This feature develops into a full peak shape anomaly for stable pyrochlore compounds 1.5 $\leq$ $\textit{x}$ $\leq$ 1.99 (Fig. 5c). For the biphasic region (0.5 $<$ $\textit{x}$ $<$ 1.5), we observed only a single broad peak which might consist of the peak anomaly from both the ordered phase. The study on the dilution of Dy$_{2}$Ti$_{2}$O$_{7}$ by Y and Lu atom by Snyder $\textit{et al.}$ \cite{snyder2004quantum} shows that the evolution of freezing temperature is independent of structural symmetry (Dy$_{2-\textit{x}}$B$_{\textit{x}}$Ti$_{2}$O$_{7}$: B = Y, Lu form stable pyrochlore structure for all value of \textit{x}). The effect of Ca$^{2+}$ substitution on the spin freezing in Dy$_{2}$Ti$_{2}$O$_{7}$ is similar to the Y$^{3+}$, however, it suppresses the freezing anomaly to a greater extent \cite{snyder2001spin,anand2015investigations}.
	
	\begin{figure}[htbp]
	\begin{center}
		\vspace{2pt}
		\includegraphics[width=8cm,height=7cm]{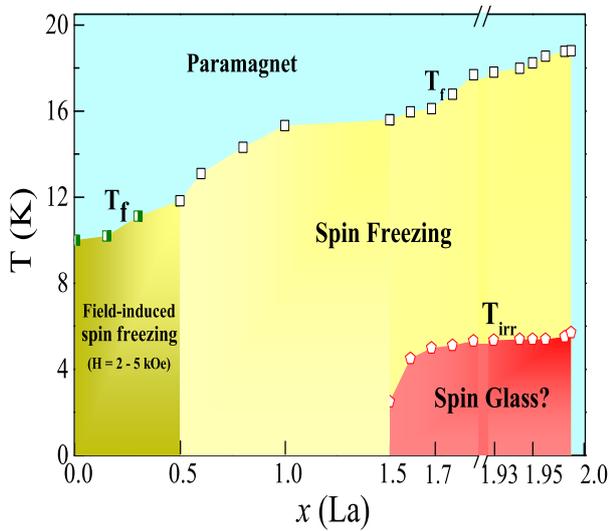}
		\caption{Magnetic phase diagram of Dy$_{2-\textit{x}}$La$_\textit{x}$Zr$_{2}$O$_{7}$; $\textit{x}$ = 0 - 2.0 showing the evolution of spin freezing transition on La substitution. Data points marked by star are taken from Ref[16] indicating the field induced spin ice phase for $\textit{x}$ $\leq$ 0.5. The compositions between 0.5 $<$ $\textit{x}$ $\leq$ 1.99 represents the spin freezing transition ($\textit{T}$$_{f}$) and the low temperature state culminates to glassy behavior marked by $\textit{T}$$_{irr}$.}
	\end{center}
	\vspace{-20pt}
\end{figure}
	
	The $\chi^{\prime}$(T) and $\chi^{\prime\prime}$(T) curves show a monotonous rise for all the excitation frequencies with a decrease in temperature (see supplementary Fig. S7). Though the frequency dependence of spin freezing transition indicates the glassy behavior, the obtained value of Mydosh parameter; \textit{p} = $\Delta$T$_{f}$/T$_{f}$$\Delta$(ln$\textit{f}$) $\sim$ 0.80 - 0.2 are much larger than the typical spin-glass value of $\it{p}$ $\sim$ 0.005 - 0.01 \cite{mydosh1993spin}, and thus point towards the unusual spin freezing in these systems. Therefore, an alternate explanation is required to understand the spin relaxation in these structurally clean systems. Further the frequency dependence of $\chi^{\prime}$ follows the Arrhenius Law, given by \textit{f} = \textit{f}$_{o}$exp(-\textit{E}$_{a}$/\textit{k}$_{B}$\textit{T}), where \textit{E}$_{a}$ is the average activation energy for spin fluctuations (shown in supplementary Fig. S8). The obtained values of the relaxation time ($\textit{$\tau$}$) and activation energy \textit{E}$_{a}$ are in the range of 10$^{-6}$ sec and 150 ($\pm$ 20) K respectively. The variations of \textit{E}$_{a}$ and \textit{$\tau$} with La compositions are shown in Fig. 5d, which is of the order of single-ion anisotropy energy and reasonably supports the individual spin flips in these systems. The slow relaxation of spin can be explained based on an increase in energy barrier in the substituted compounds compared to DZO \cite{ramon2019absence}. Further, it is notable that in the stable pyrochlores (\textit{x} $\geq1.5$) \textit{E}$_{a}$ substantially decreases on increasing the dilution. It is possibly due to the re-entrance symmetry of the pyrochlore structure with La substitution, which effectively changes the magnitude of crystal field spacing associated with the change in lattice constant and electronic structure. Indeed, the x-ray studies indicate that the lattice parameter increases monotonically with La substitution in consistence with the decrease in \textit{E}$_{a}$. The observed behavior is similar to the structural studies on Y and Lu doped Dy$_{2}$Ti$_{2}$O$_{7}$ where the rise in \textit{E}$_{a}$ correlates with the monotonic decrease in the lattice parameter with dilution \cite{snyder2004quantum}. \\
A more striking difference between the observed spin freezing and glassy freezing is found in the frequency distribution of the spin relaxation process. In Fig. 5e, we plot the frequency dependence of normalized $\chi^{\prime\prime}$ and analysed in term of a Casimir-du Pr$\acute{e}$ relation. This predicts that for a individual spin relaxation time $\textit{$\tau$}$, $\chi^{\prime\prime}$($\it{f}$) = $\it{f}$$\textit{$\tau$}$[($\chi_{T}$ - $\chi_{S}$)/(1 + $\it{f}^{2}$$\tau^{2}$)] where $\chi_{T}$ and $\chi_{S}$ are the isothermal and adiabatic susceptibility respectively \cite{anand2015investigations}. We find the close proximity of our data to the theoretical results corresponding to single spin relaxation. The presence of single spin relaxation is also evident from the Cole-Cole plot (see supplementary Fig. S8) \cite{anand2015investigations}. The data points on $\chi^{\prime\prime}$($\chi^{\prime}$) curves corresponding to different frequencies fall on the semi-circular arc of varying diameters for all compositions. The inset of Fig. 5d shows the variation in $\alpha$ (width of relaxation times) for different compositions. The obtained value of $\alpha$ is quite large from other dense magnetic systems (DyP$_{1-\textit{x}}$V$_{\textit{x}}$O$_{4}$, Rb$_{2}$Cu$_{1-\textit{x}}$Co$_{\textit{x}}$F$_{4}$) exhibiting glassy behavior in which the relaxation times are typical of several orders \cite{dirkmaat1987frequency,dekker1989activated}. The shape of arc remains unchanged upon La substitution and indicates that the distribution of relaxation times is unaffected and follows the single spin relaxation process for Dy$_{2-\textit{x}}$La$_\textit{x}$Zr$_{2}$O$_{7}$. 

We further measured the frequency (1 $\leq$ \textit{f} $<$ 1 kHz) dependence of $\chi_{ac}$ to estimate the relaxation time ($\textit{$\tau$}$) of the spin freezing (see supplementary Fig. S9). The $\chi^{\prime\prime}$(\textit{f}) shows a well-defined peak above $\textit{T}$ = 10 K, which shifts with temperature. It is to mention that the evolution of $\textit{$\tau$}$ for Dy$_{2}$Ti$_{2}$O$_{7}$ is linked with the crossover from thermal to quantum spin relaxation \cite{snyder2003quantum,snyder2004low}. Unlike Dy$_{2}$Ti$_{2}$O$_{7}$, the obtained values of $\textit{$\tau$}$ for Dy$_{2-\textit{x}}$La$_\textit{x}$Zr$_{2}$O$_{7}$ ($\textit{x}$ = 1.5 - 2.0) follow the Arrhenius behavior. This activated relaxation is responsible for the high-temperature spin freezing observed at \textit{T}$_{f}$ $\sim$ 17.5 K. These data show that $\textit{$\tau$(x)}$ rises monotonically on increasing non-magnetic dilution. It is to note that DZO shows a slower spins relaxation ($\tau$ = 10$^{-1}$ s) at the lowest measuring temperature in comparison to Dy$_{2}$Ti$_{2}$O$_{7}$ ($\tau$ = 10$^{-3}$ s) \cite{snyder2004low}. For $\textit{x}$ = 1.0, no relaxation peak was observed in $\chi^{\prime\prime}$($\it{f}$) plot, but a very weak signal of freezing transition was detected in $\chi^{\prime\prime}$(\textit{T}) plot. In Y and Lu doped Dy$_{2}$Ti$_{2}$O$_{7}$ \cite{snyder2004quantum}, the unusual re-entrance of the high-temperature spin-freezing can be understood from the spin relaxation time. The non-monotonic nature of \textit{$\tau$(x)} on the La concentration for Dy$_{2-\textit{x}}$La$_\textit{x}$Zr$_{2}$O$_{7}$ is associated with the increase in energy barrier with dilution. Our results for La substitution from 0 to 1.99 showed the disappearance of freezing transition in the disordered pyrochlore phase and its re-emergance on achieving the structural stability in the absence of magnetic field. 
	
	The magnetic phase diagram of Dy$_{2-\textit{x}}$La$_\textit{x}$Zr$_{2}$O$_{7}$; $\textit{x}$ = 0 - 2.0 has been drawn in Fig. 6. The label \textit{T}$_{f}$ and \textit{T}$_{irr}$ were extracted from the ac and dc magnetic susceptibility. It is clearly seen that the increase in La concentration stabilizes the Dy$_{2}$Ti$_{2}$O$_{7}$ like spin freezing transition for $\textit{x}$ = 0.5 - 1.99. The structurally biphasic region (0.5 $<$ \textit{x} $<$ 1.5) shows only one type of spin freezing only. It quite possible that the spin freezing corresponding to the minor phase get overwhelmed by the spin freezing of majority phase, as the spin freezing transitions of these phases are quite close to each other. For the compositions \textit{x} $\geq$ 1.5, a new spin glass like freezing is observed at low temperatures. The end member of the series La$_{2}$Zr$_{2}$O$_{7}$ does not show any sort of the spin freezing. 
	
	In conclusion, we have shown that the position of the oxygen atom plays a significant role in determining the crystal structure of Dy$_{2-\textit{x}}$La$_\textit{x}$Zr$_{2}$O$_{7}$ and the pyrochlore phase gets stabilized on La substitution. The low-temperature magnetic ground state of the system evolves from a field-induced spin freezing state for La compositions of 0 $\leq$ $\textit{x}$ $\leq$ 0.3 to the spin freezing in the absence of magnetic filed for 1.5 $\leq$ $\textit{x}$ $\leq$ 1.99. The spin dynamics of the system suggest slower spin relaxation in comparison to Dy$_{2}$Ti$_{2}$O$_{7}$. For 1.5 $\leq$ $\textit{x}$ $\leq$ 1.99, the spin freezing is followed by a spin glass	like state below 6 K. These study suggest the robustness of the spin freezing state in Dy pyrochlore system and may have significant bearing on the observed spin ice behavior in Dy$_{2}$Ti$_{2}$O$_{7}$. Our studies would be useful in the understanding of the high temperature spin freezing state in the spin ice systems.  
	
	We acknowledge the Advanced Material Research Center, IIT Mandi, for the experimental facility. Sheetal is thankful to IIT Mandi and MHRD India for the Senior Research Fellowship.
	
	\bibliography{Ref}
		
	\section{Supplementary information}	
	
	\subsection{Experimental Details}
	
	The polycrystalline compounds of Dy$_{2-x}$La$_{x}$Zr$_{2}$O$_{7}$ (x: 0.5, 0.6, 0.8, 1.0, 1.5, 1.6, 1.7, 1.8, 1.9, 2.0) were prepared by the standard solid-state reaction method  \cite{devi2020emergence}. The constituent oxides: Dy$_{2}$O$_{3}$ (Sigma Aldrich, $\geq$99.99$\%$ purity), La$_{2}$O$_{3}$ (Sigma Aldrich, $\geq$99.999$\%$ purity) and ZrO$_{2}$ (Sigma Aldrich, 99$\%$ purity) were reacted in the alumina crucible at 1350$^{o}$C in ambient atmosphere for 50 hours. The reaction at this temperature was done thrice with the intermediate grindings. The obtained compounds were further pelletized and sintered at 1350$^{o}$C for 50 hours. The other compounds \textit{x} = 1.93, 1.95, 1.96, 1.97, 1.985, 1.99 are prepared with solid-state reaction method with slight change in the annealing condition. In the dilution regime (1.5 $\leq$ \textit{x} $\leq$ 1.99) we observed consistency in the structural and magnetic parameter. The spin freezing and irreversibility in dc magnetization is remain prominent even up to the dilution of 99.5$\%$. Magnetic and Heat capacity studies on the Dy$_{2}$Zr$_{2}$O$_{7}$ reported by Ramon  $\textit{et al.}$ matches well with our previous studies \cite{ramon2019absence}. Therefore, we believe that the reported results in this manuscript are intrinsic to these system, and are not affected by the preparation conditions. Additionally, we have performed all magnetization measurements on the same piece of sample.
	
	The powder X-ray diffraction measurement on all the compounds was performed using the Rigaku x-ray diffractometer in the 2$\theta$ range of 10 - 90$^{o}$ with the step size of 0.02$^{o}$. The Rietveld refinement of the x-ray diffraction (XRD) pattern was performed using Fullprof Suite software. Raman spectra of the compound were obtained at 300 K in backscattering geometry by using a Horiba HR-Evolution spectrometer with a 532 nm excitation laser. We study the magnetization as well as the real ($\chi^{\prime}$) and imaginary parts ($\chi^{\prime\prime}$) of the ac susceptibility ($\chi_{ac}$). The magnetic measurements were performed using Quantum Design built Magnetic Property Measurement System (MPMS) above 1.8 K. In thermoremanent magnetization (TRM) the sample was cooled from 10 K to the desired temperature in the presence of magnetic field, reducing the field to zero at a rate of 0.1 T/min, and then measuring the dc magnetization as a function of time. The isothermal magnetization (IRM) was collected by cooling the sample from 10 K in the absence of a field, then cycling the field from 0$\rightarrow$\textit{H}$\rightarrow$0 and measuring the magnetization as a function of time. The sample was held at H for at least 2 hours to obtain nearly complete relaxation in the field. 
	
\subsection{Structural studies}

\begin{figure}[htbp]
	\begin{center}
		\includegraphics[width=8cm, height=11cm]{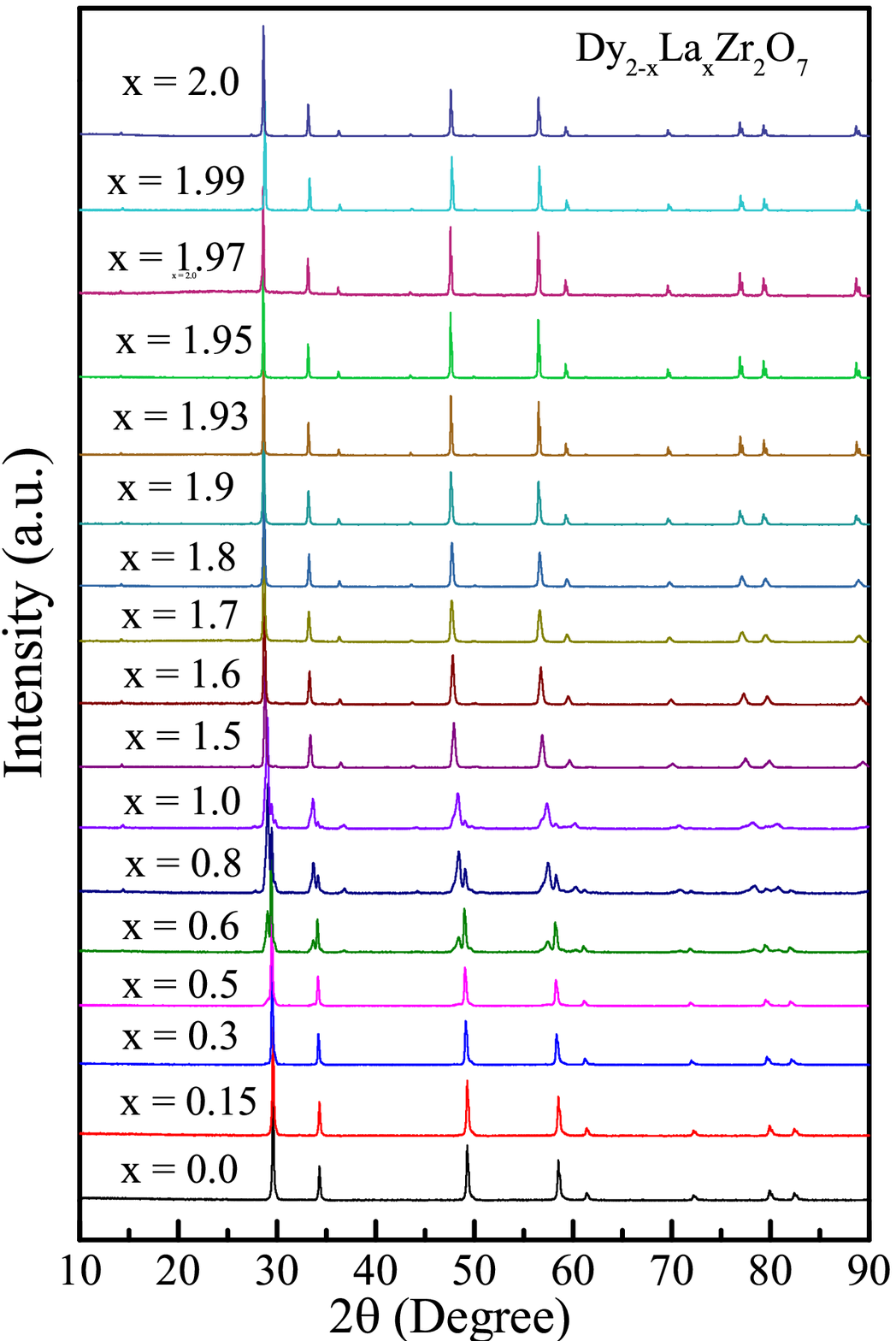}
		\vspace{-5pt}
	\end{center}
	\textbf{Figure S1}: Room temperature x-ray diffraction pattern of Dy$_{2-x}$La$_{x}$Zr$_{2}$O$_{7}$ (\textit{x} = 0.0 - 2.0). A clear shift in peak positions towards lower $\theta$ values indicates the increase in lattice constant with increase in the La concentration.
\end{figure}

\begin{figure}[htbp]
	\begin{center}
		\includegraphics[width=8.5 cm, height= 14 cm]{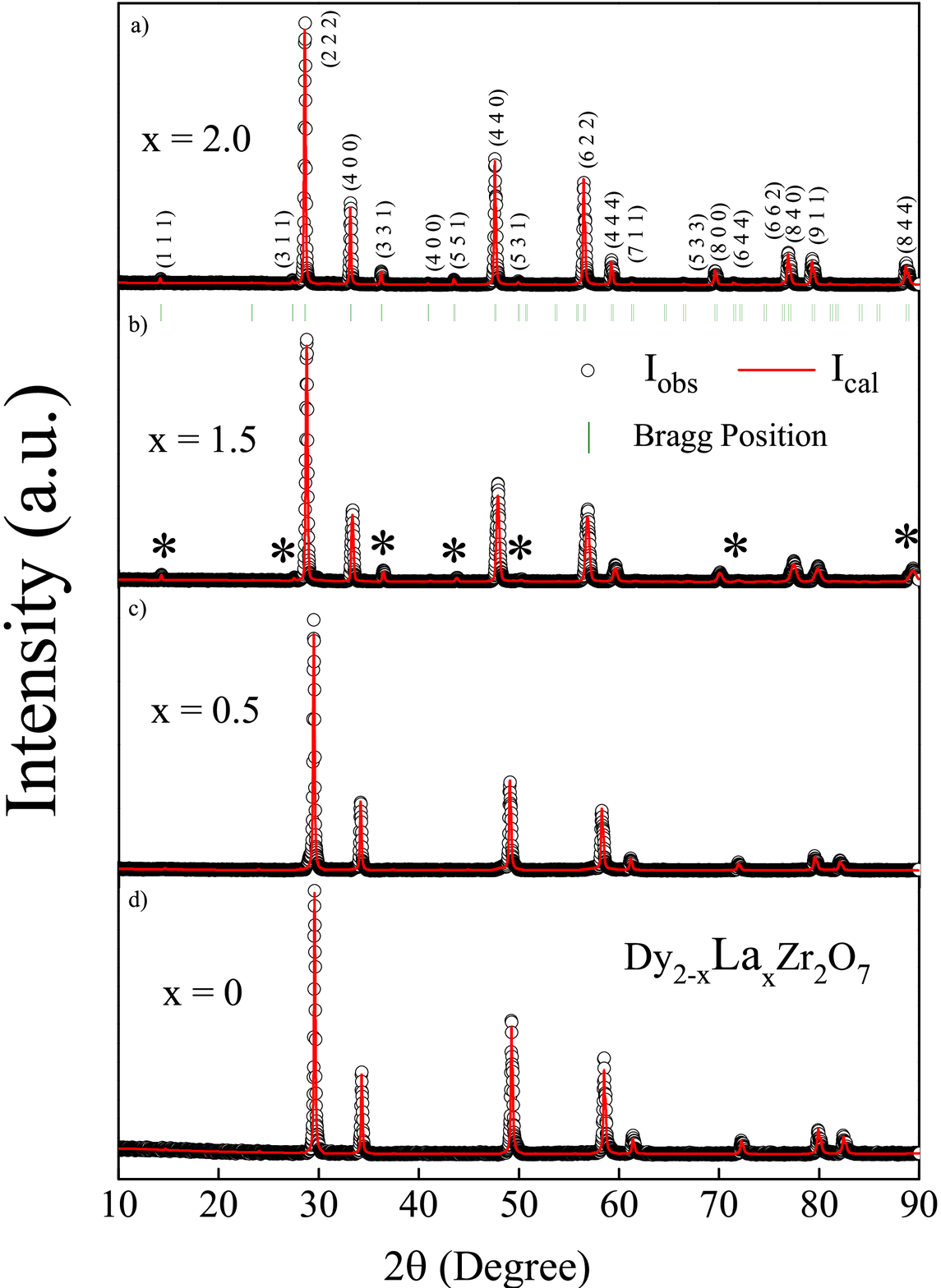}
		\vspace{5pt}
	\end{center}
	\textbf{Figure S2}: The Rietveld refined x-ray diffraction pattern of Dy$_{2-x}$La$_{x}$Zr$_{2}$O$_{7}$ (\textit{x} = 0, 0.5, 1.5, 2.0) using Fd$\bar{3}$\textit{m} space group. Peaks marked by star are the superstructure peaks belonging to pyrochlore structure, and indicates the pyrochlore type ordering in the compounds.
	
\end{figure}

Figure S1 and S2 show the x-ray diffraction pattern of Dy$_{2-x}$La$_{x}$Zr$_{2}$O$_{7}$ (\textit{x} = 0 - 2). The XRD patterns of the compounds are fitted nicely in Fd$\bar{3}$ space group and show the single phase of these compounds. Structural evolution from weak pyrochlore phase (0 $\leq$ \textit{x} $\leq$ 0.5) to completely stabilised pyrochlore phase (1.5 $\leq$ \textit{x} $\leq$ 2) is clear from the x-ray studies. The observation of the superstructure peaks for \textit{x} $\geq$ 0.5 directs the structural geometry towards the stable pyrochlore phase. The refined parameters for all the compositions are listed in Table I. Fig. 3 shows the refined x-ray pattern of a mixed-phase compound DyLaZr$_{2}$O$_{7}$, where XRD data is nicely fitted with the parameters of disordered fluorite and pyrochlore phase using the same space group and three different lattice parameters. Though the majority phase ($\sim$ 97$\%$) is pyrochlore (a = 10.6669 \AA), there are two other $\sim$ 2$\%$ and $\sim$ 1$\%$ weak pyrochlore phases with lattice parameters a = 10.4496(3) \AA and a = 10.3737(3) \AA.	

\begin{figure}[htbp]
	\begin{center}
		\includegraphics[scale = 0.45]{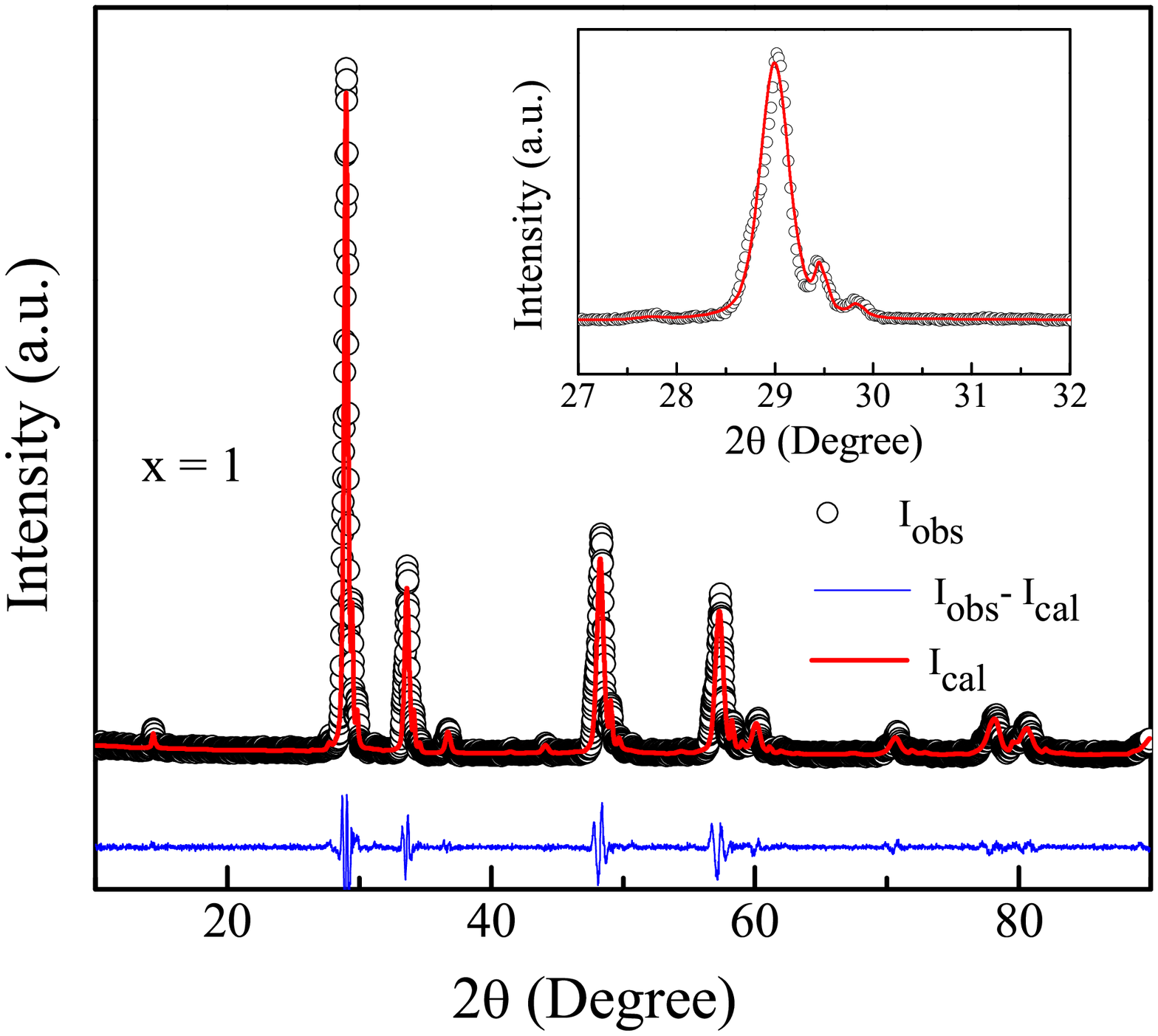}
		\vspace{-5pt}
	\end{center}
	\textbf{Figure S3}: The Rietveld refined x-ray diffraction pattern of DyLaZr$_{2}$O$_{7}$. Inset shows the splitting of main peak and refinement with disordered fluorite and pyrochlore phase more clearly.
\end{figure}  

\begin{figure}[ht]
	\begin{center}
		\includegraphics[width=8.5cm,height=3.5cm]{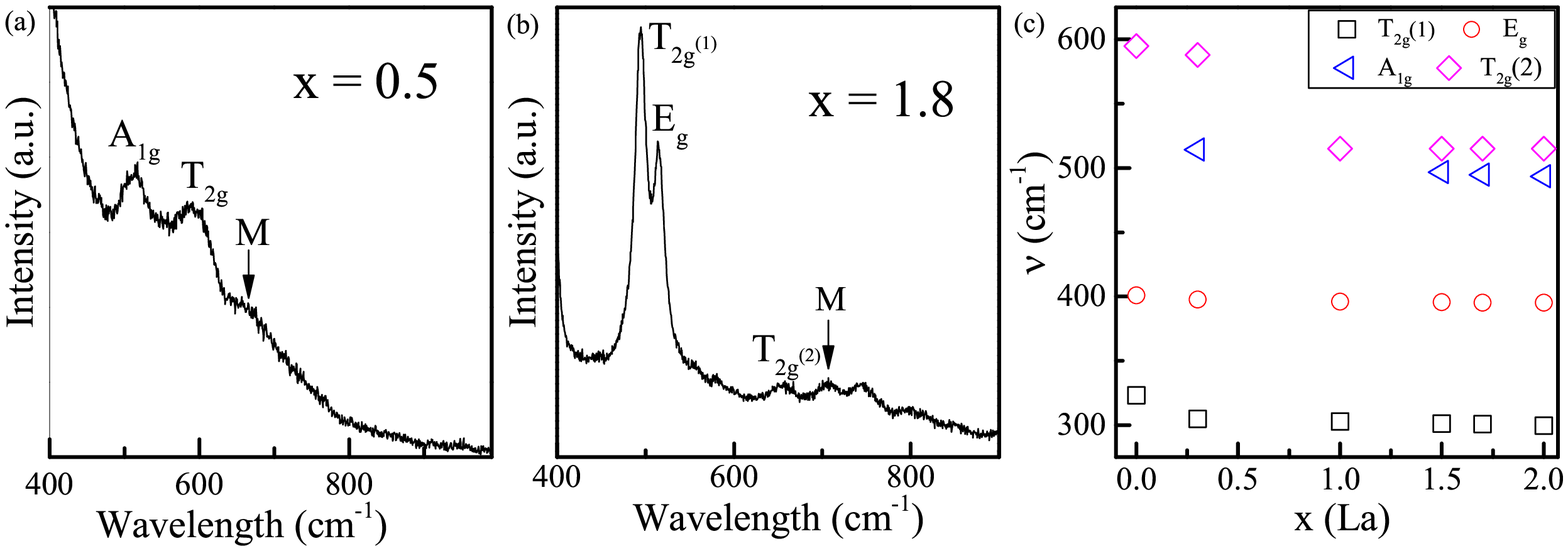}
		\vspace{-10pt}
	\end{center}
	\textbf{Figure S4}: (a $\&$ b) Room temperature Raman spectra of Dy$_{2-x}$La$_{x}$Zr$_{2}$O$_{7}$; (0.5, 1.8) showing the presence of Raman modes corresponding to pyrochlore structure. (c) Variation in the phonon frequency of Raman modes with increase in La concentration.
\end{figure}

\pagebreak

\subsection{Magnetic studies}

\begin{figure}[htbp]
	\begin{center}
		\includegraphics[width=8.5cm,height=11cm]{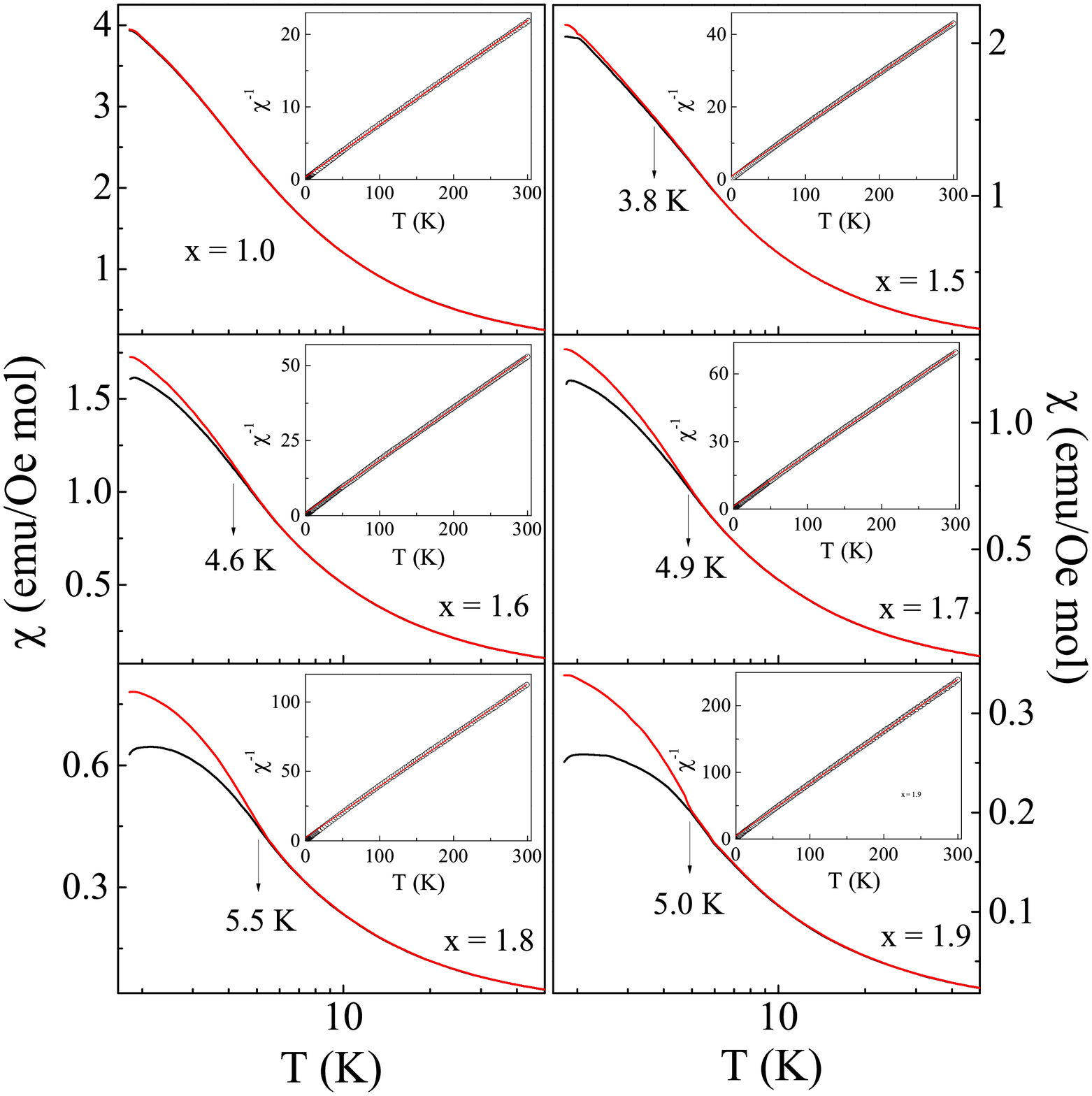}
		\vspace{-10pt}
	\end{center}
	\textbf{Figure S5a}: Dc magnetic susceptibility $\chi_{dc}$ (ZFC and FC) versus \textit{T} at \textit{H} = 100 Oe for Dy$_{2-\textit{x}}$La$_{\textit{x}}$Zr$_{2}$O$_{7}$ (0 $\leq$ \textit{x} $\leq$ 2). Inset: Curie-Weiss fit of the magnetization data at 100 Oe in the temperature range 30 - 300 K.
\end{figure}

\begin{figure}[htbp]
	\begin{center}
		\includegraphics[width=9cm,height=6cm]{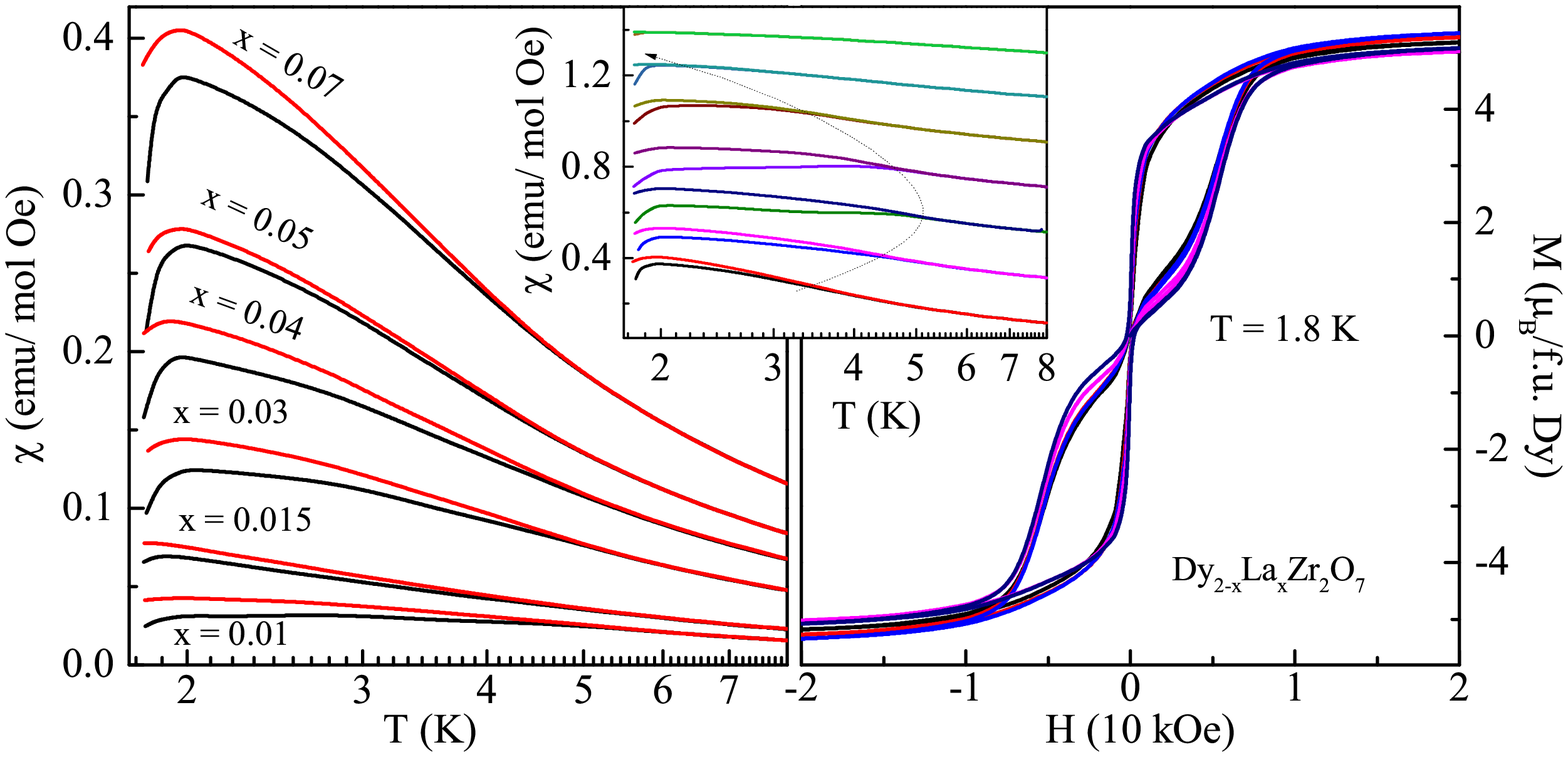}
		\vspace{-10pt}
	\end{center}
	\textbf{Figure S5b}: (Left) Dc magnetic susceptibility $\chi_{dc}$ (ZFC and FC) versus \textit{T} at \textit{H} = 100 Oe and (Right) isothermal magnetization for Dy$_{2-\textit{x}}$La$_{\textit{x}}$Zr$_{2}$O$_{7}$ (1.93 $\leq$ \textit{x} $\leq$ 1.99). Inset: ZFC and FC data of \textit{x} = 1.93 to show the field dependence of \textit{T}$_{irr}$.
\end{figure}

\begin{figure}[htbp]
	\begin{center}
		\includegraphics[width=9cm,height=7cm]{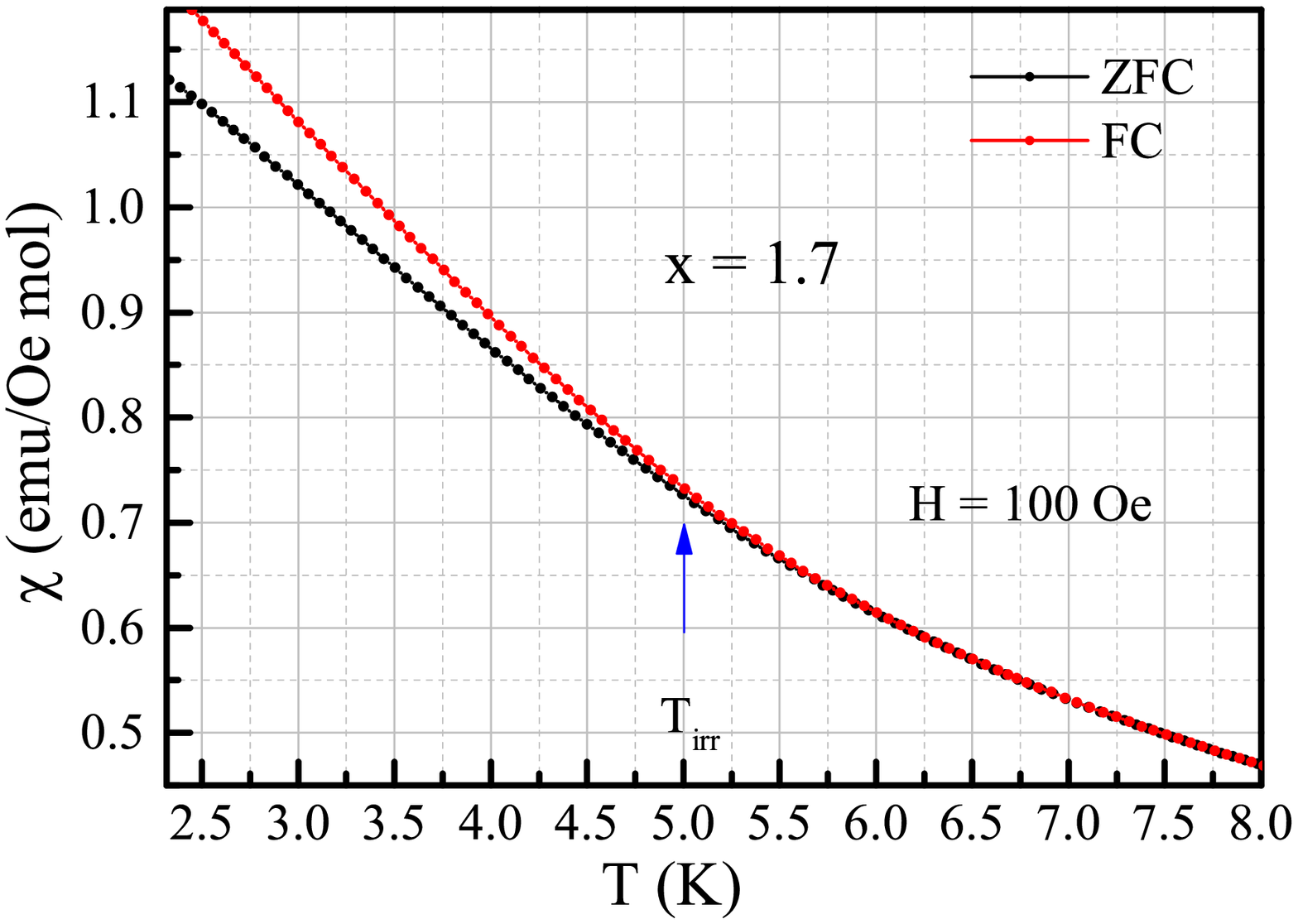}
		\vspace{-10pt}
	\end{center}
	\textbf{Figure S5c}: Temperature dependence of magnetization data for x = 1.7 composition. The point of visible separation between zfc and fc curve is identified as \textit{T}$_{irr}$. However there can be an error of $\pm$ 0.2 K in \textit{T}$_{irr}$ for these compounds. 
\end{figure}

\begin{figure}[htbp]
	\begin{center}
		\includegraphics[width=8cm,height=10cm]{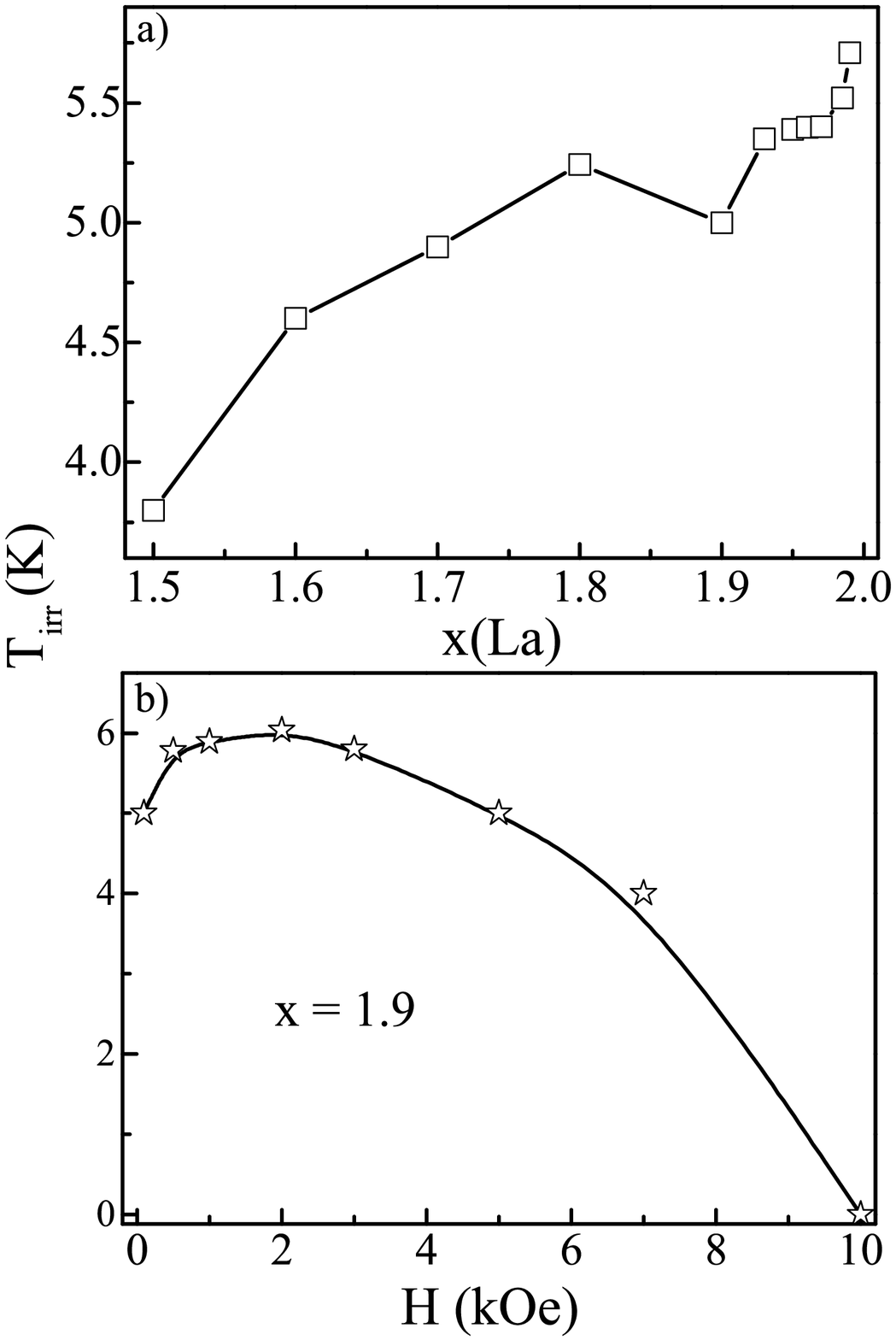}
		\vspace{-10pt}
	\end{center}
	\textbf{Figure S6}: (a) La concentration dependence of \textit{T}$_{irr}$ taken as a bifurcation point of ZFC and FC data in Dy$_{2-\textit{x}}$La$_{\textit{x}}$Zr$_{2}$O$_{7}$ (1.5 $\leq$ \textit{x} $\leq$ 1.99). (b) Field dependence of \textit{T}$_{irr}$ measured for Dy$_{0.1}$La$_{1.9}$Zr$_{2}$O$_{7}$. 
\end{figure}

Figures S5a and S5b shows the dc magnetization data of Dy$_{2-\textit{x}}$La$_{\textit{x}}$Zr$_{2}$O$_{7}$ (1.0 $\leq$ \textit{x} $\leq$ 1.99) and the inset of each plot shows the CW fit of the inverse susceptibility data. The inverse susceptibility data of Dy$_{2-\textit{x}}$La$_{x}$Zr$_{2}$O$_{7}$ (measured at \textit{H} = 100 Oe) is fitted with Curie-Weiss law in the high temperature regime (\textit{T} $\geq$ 30 K) using the relation $\chi$ = \textit{C}/(\textit{T} - $\theta_{cw}$), where \textit{C} and $\theta_{cw}$ stands for Curie constant and Curie-Weiss temperature respectively. From the Curie constant, the effective magnetic moment was calculated using $\mu_{eff}$ = $\sqrt{3\textit{k}_{B}\textit{C}/\textit{N}_{A}}$, where \textit{N}$_{A}$ is the Avogadro$^{\prime}$s number. The obtained value of $\theta_{cw}$ and $\mu_{eff}$ for all the substituted compounds are listed in Supplementary Table II. The negative value of $\theta_{cw}$ signifies the dominance of AFM interactions in the system.  Fig. S6 shows the variation in the bifurcation point (\textit{T}$_{irr}$) with La substitution and the magnetic field. Fig. S7 shows the temperature dependence of real and imaginary parts of ac susceptibility measured at various frequencies between 10 - 1000 Hz. A clear relaxation peak starts evolving on entering the biphasic region and becomes prominent in the stable pyrochlore phase. In contrast, parent compound Dy$_{2}$Zr$_{2}$O$_{7}$ exhibits a different behavior and paramagnetic down to 1.8 K. The magnetic phase transition is consistent with the structural phase transition (discussed in the main text). The ac susceptibility data is analyzed by Cole-Cole plot (Fig. S8 (left)) and Arrhenius fit (Fig. S8 (right)), and the obtained parameters are plotted in the main text. To further analyze the relaxation behavior in detail, ac susceptibility measurements were performed as a function of frequencies for \textit{T} $<$ 40 K and are shown in Fig. S9.

\begin{table}[htbp]
	\caption{Magnetic parameters of Dy$_{2-\textit{x}}$La$_{\textit{x}}$Zr$_{2}$O$_{7}$; 0.0 $\leq$ \textit{x} $<$ 2.0}
	\begin{tabular}{ccccccccc}
		\hline
		x(La) && $\mu_{eff}$ ($\mu_{B}$) && $\theta_{cw}$ (K) &&& $\it{p}$ & Ref\\
		\hline
		0     && 16.4 (1.09)  && -9.6(2)  &&& - & \cite{devi2020emergence} \\
		0.3   && 14.12 (0.65) && -7.14(1) &&& - &   ,, \\
		0.5   && 12.94 (0.43) && -7.02(3) &&& - &  This study \\
		1.0   && 10.51 (0.35) && -5.63(2) &&& - &   ,, \\
		1.5   && 7.56 (0.26)  && -6.61(2) &&& 0.80 &   ,, \\
		1.6   && 6.80 (0.23)  && -5.64(2) &&& 0.71 &   ,, \\
		1.7   && 5.93 (0.19)  && -5.80(2) &&& 0.66 &   ,, \\
		1.8   && 4.67 (0.18)  && -6.63(7) &&& 0.64 &   ,, \\
		1.9   && 3.21 (0.09)  && -6.17(5) &&& 0.6  &   ,, \\
		1.93  && 2.83 (0.05)  && -6.11(3) &&& 0.25 &   ,, \\ 
		1.95  && 2.39 (0.03)  && -6.05(1) &&& 0.26 &   ,, \\
		1.96  && 2.22 (0.025)  && -6.21(3) &&& 0.27 &   ,, \\    
		1.97  && 1.80 (0.017)  && -4.41(3) &&& 0.26 &   ,, \\
		1.985 && 1.25 (0.013)  &&  2.89(2) &&& 0.28 &   ,, \\ 
		1.99  && 0.84 (0.009)  &&  2.81(6) &&& 0.26 &   ,, \\
		\hline
	\end{tabular} 
\end{table} 

\pagebreak	

\begin{figure*}[ht]
	\begin{center}
		\vspace{30pt}
		\includegraphics[width= 12cm,height=14 cm]{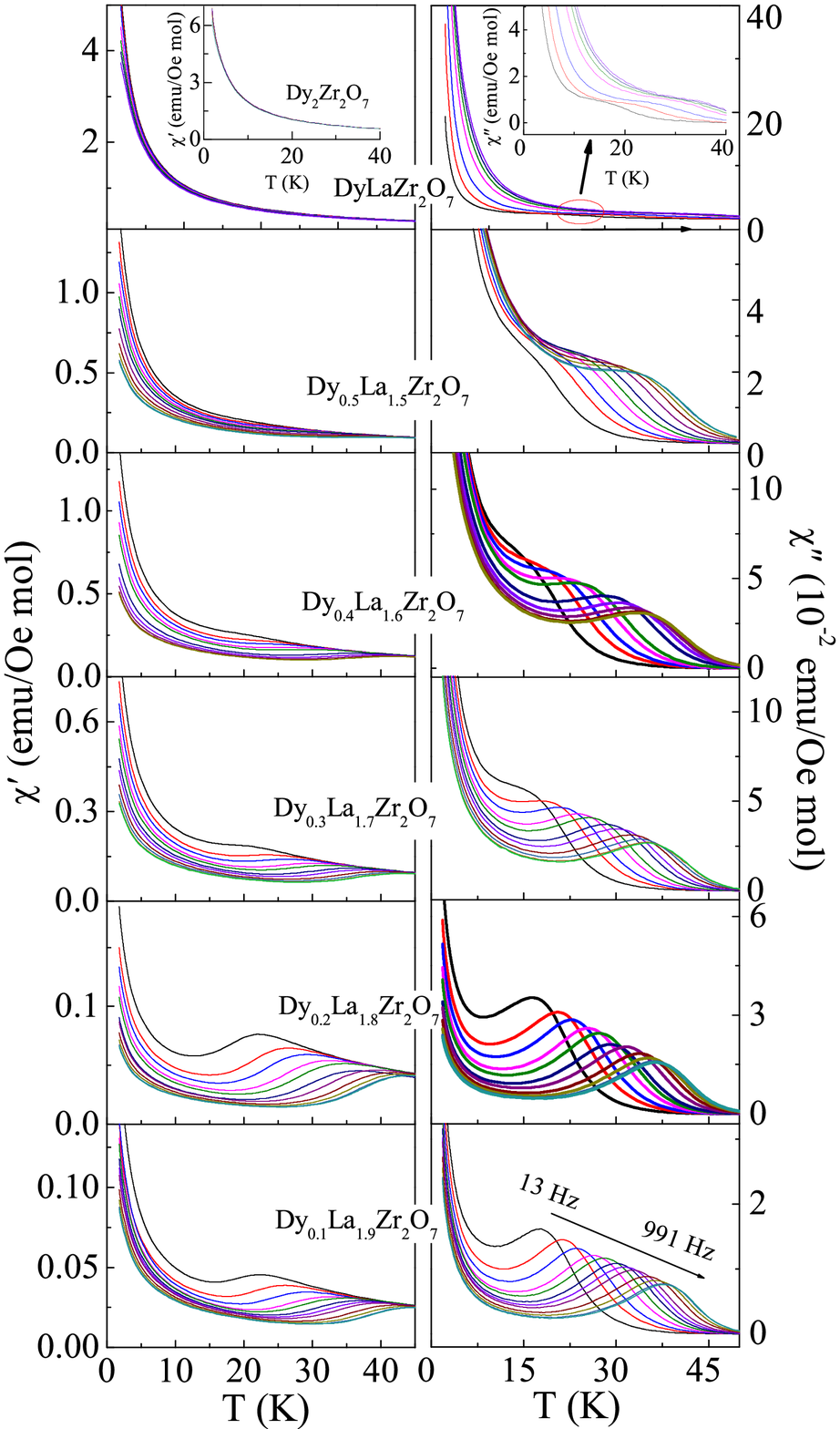}
		\vspace{2pt}
	\end{center} 
	\textbf{Figure S7}: Temperature dependence of real and imaginary part of ac susceptibility for Dy$_{2-\textit{x}}$La$_{\textit{x}}$Zr$_{2}$O$_{7}$ (0 $\leq$ \textit{x} $\leq$ 1.9) at various frequencies ranges between 10 - 1000 Hz. Left inset shows the temperature response of $\chi^{\prime}$ of Dy$_{2}$Zr$_{2}$O$_{7}$ and right inset shows the temperature response of $\chi^{\prime\prime}$ for DyLaZr$_{2}$O$_{7}$.
\end{figure*}

\begin{figure*}[htbp]
	\begin{center}
		\includegraphics[width= 10 cm, height= 12 cm]{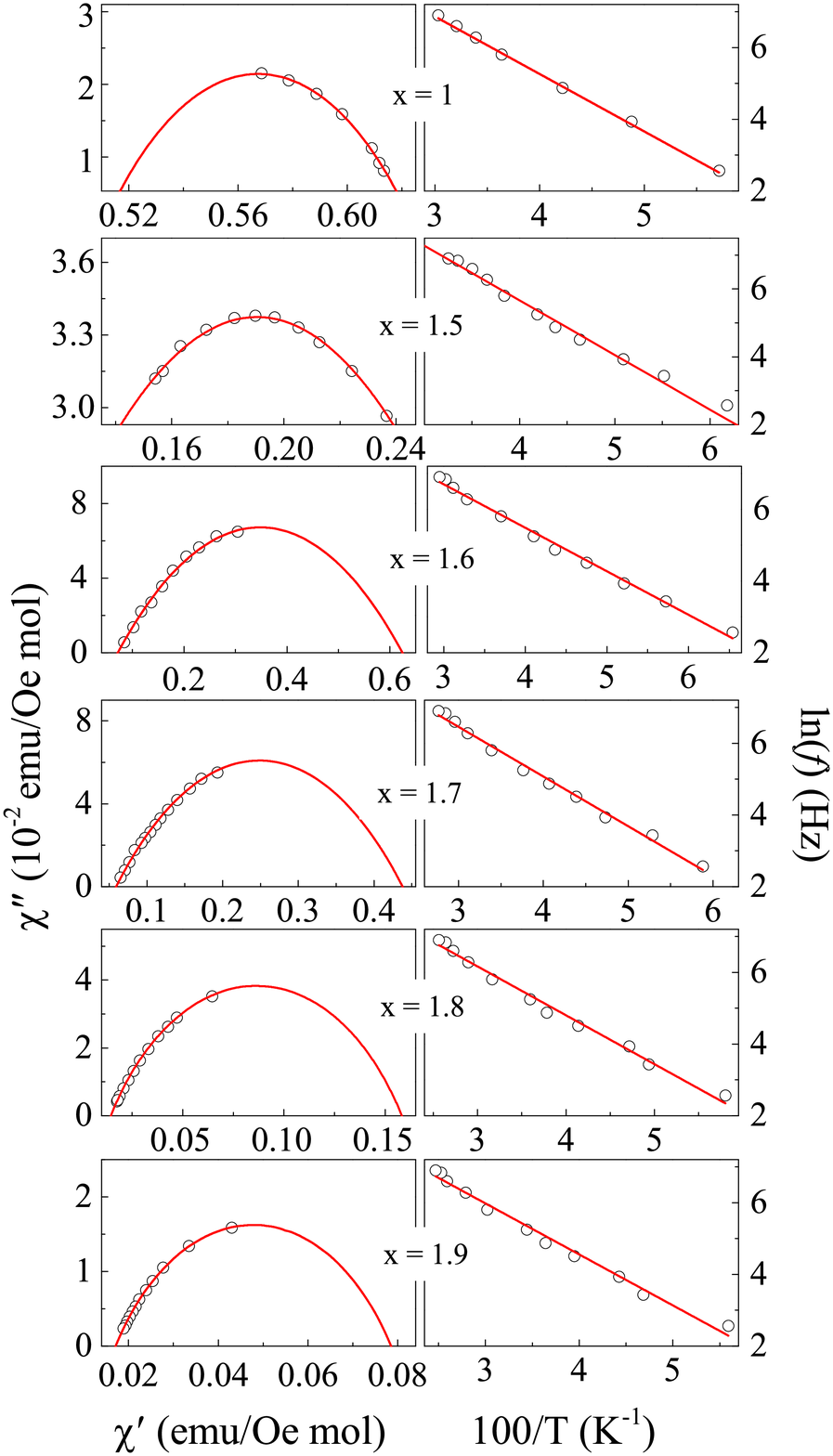}
	\end{center}
	\vspace{2pt}
	\textbf{Figure S8}: Cole-Cole plot of $\chi^{\prime}$ and $\chi^{\prime\prime}$ data around the freezing temperature and Arrhenius fit of Dy$_{2-\textit{x}}$La$_{\textit{x}}$Zr$_{2}$O$_{7}$ (1.5 $\leq$ \textit{x} $\leq$ 1.99).
\end{figure*}

\begin{figure*}[htbp]
	\begin{center}
		\vspace{1pt}
		\includegraphics[width= 12 cm, height= 13 cm]{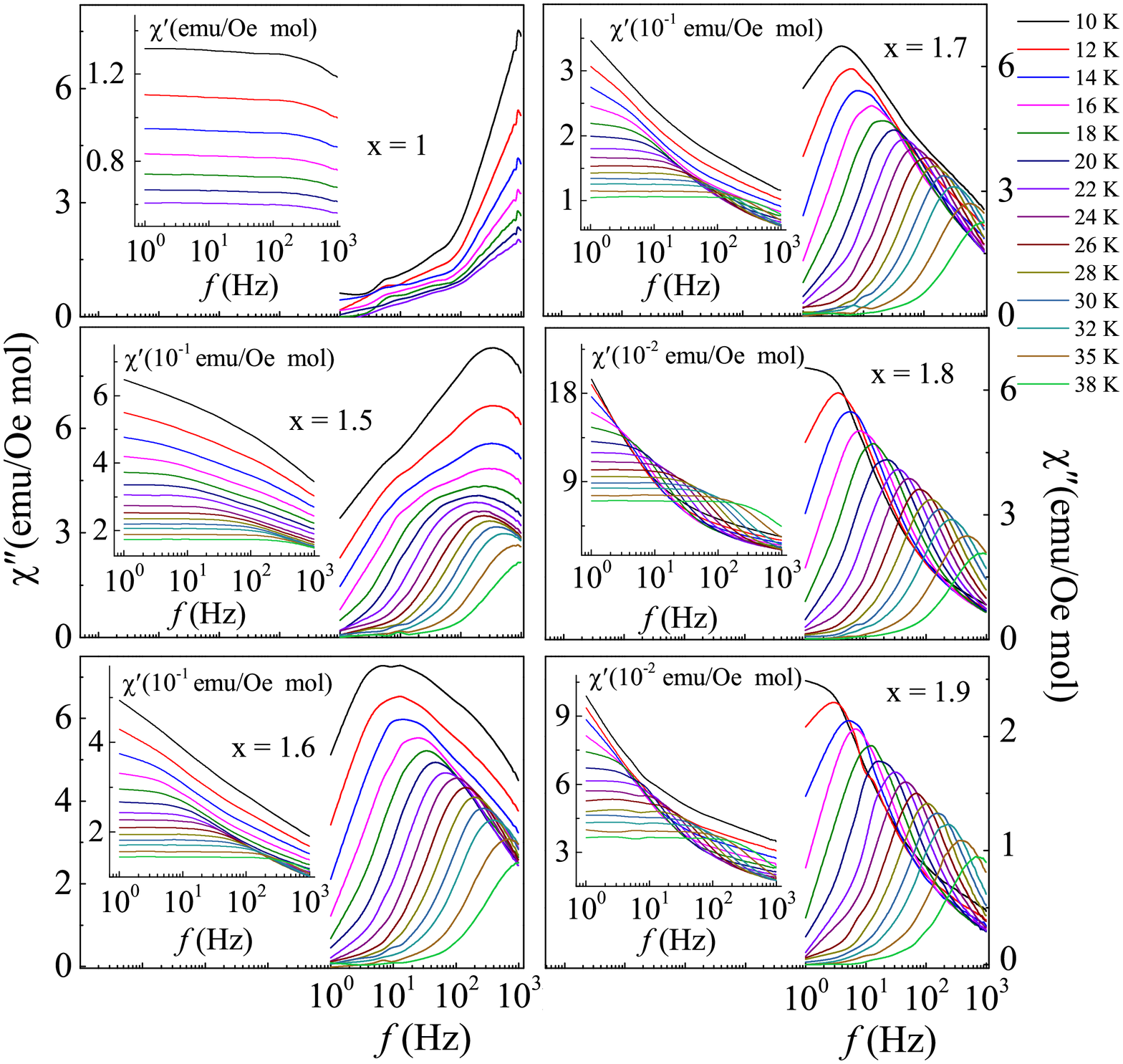}
		\vspace{5pt}
	\end{center}
	\textbf{Figure S9}: Frequency dependence of the imaginary part of ac susceptibility of Dy$_{2-\textit{x}}$La$_{\textit{x}}$Zr$_{2}$O$_{7}$ (1.0 $\leq$ \textit{x} $\leq$ 1.9) below and above the freezing temperature in zero applied field. The presence of prominent single peak is the signature of single characteristic relaxation time.
\end{figure*}

\pagebreak

\begin{table*}[htbp]
	\caption{Structural parameters Dy$_{2-\textit{x}}$La$_{\textit{x}}$Zr$_{2}$O$_{7}$; 0.0 $\leq$ \textit{x} $\leq$ 2.0}
	\begin{tabular}{ccccccccccccccccccccccccccccccccccccccccccc}
		\hline
		x(La) && r$_{A}$/r$_{B}$&& $\chi^{2}$ && x(O) && a($\AA$)&& Phase$\%$ &&Ref \\
		\hline
		0.0  && 1.43 && 1.46 && 0.3667(23) && 10.4511(3) && 100 &&\cite{devi2020emergence} \\
		0.15 && 1.44 && 2.26 && 0.3682(28) && 10.4832(4) && 100 &&  ,, \\
		0.3  && 1.45 && 2.22 && 0.3694(29) && 10.4972(3) && 100 &&  ,, \\
		0.5  && 1.47 && 1.51 && 0.3710(2) && 10.5521(6) && 100 &&This study\\
		\hline
		0.6  && 1.48 && 2.15 && 0.3271(13) && 10.5994(4) && 60 &&  ,,\\
		&&      &&      && 0.3729(3) && 10.4797(3) && 35 &&  ,,\\
		&&      &&      && 0.3492(2) && 10.3895(6) &&  5 &&  ,,\\
		\hline
		0.8  && 1.49 && 1.95 && 0.3273(10) && 10.6482(5) && 88 &&  ,,\\
		&&      &&      && 0.3604(3) && 10.4992(2) &&  8 &&  ,,\\
		&&      &&      && 0.3724(3) && 10.3889(8) &&  4 &&  ,,\\
		\hline
		1.0  && 1.52 && 2.85 && 0.3248(7) && 10.6669(3) &&  97 && ,,\\
		&&      &&      && 0.3630(4) && 10.4992(4) &&   2 && ,,\\
		&&      &&      && 0.3730(4) && 10.3735(2) &&   1 && ,,\\
		\hline
		1.5  && 1.56 && 1.70 && 0.3291(4) && 10.7445(3) && 100 && ,,\\
		1.6  && 1.57 && 1.36 && 0.3205(5) && 10.7633(3) && 100 && ,,\\
		1.7  && 1.58 && 1.43 && 0.3287(5) && 10.7786(3) && 100 && ,,\\
		1.8  && 1.59 && 1.23 && 0.3281(4) && 10.7812(1) && 100 && ,,\\
		1.9  && 1.60 && 1.49 && 0.3285(5) && 10.7903(1) && 100 && ,,\\
		1.93 && 1.604 && 1.57 && 0.3287(4) && 10.7958(3) && 100 && ,,\\
		1.95 && 1.606 && 1.63 && 0.3282(7) && 10.7976(1) && 100 && ,,\\
		1.96 && 1.607 && 1.55 && 0.3287(8) && 10.7981(2) && 100 && ,,\\
		1.97 && 1.608 && 1.61 && 0.3295(6) && 10.7983(1) && 100 && ,,\\
		1.985&& 1.609 && 1.52 && 0.3282(7) && 10.7989(2) && 100 && ,,\\
		1.99 && 1.610 && 1.46 && 0.3282(4) && 10.8005(2) && 100 && ,,\\
		2.0  && 1.611 && 1.40 && 0.3277(5)  && 10.8008(1) && 100 && ,,\\
		\hline
	\end{tabular}
\end{table*}

\end{document}